\documentclass[a4paper,11pt]{article}
\usepackage{amssymb,amsmath,amsfonts,makeidx,placeins,pbox,multirow}
\usepackage{graphicx,rotate,subcaption,color,slashed,cite,caption,epstopdf,verbatim}
\usepackage{longtable,tabu}
\usepackage{array}
\usepackage{pstricks}
\usepackage{color}
\usepackage{amsmath}
\usepackage{mathtools}
\usepackage{amssymb}
\usepackage[normalem]{ulem}
\usepackage{relsize}
\usepackage{pifont}% http://ctan.org/pkg/pifont
\usepackage{color}
\usepackage[normalem]{ulem}
\usepackage[colorlinks=true,
            linkcolor=magenta,
            urlcolor=blue,
            citecolor=blue]{hyperref}
\newcommand{\cmark}{\ding{51}}%
\newcommand{\xmark}{\ding{55}}%
\newcolumntype{P}[1]{>{\centering\arraybackslash}p{#1}}
\newcommand\Zdis{Z_{\rm dis}}
\newcommand\Zexc{Z_{\rm exc}}
\newcommand\missET{E_T^{\rm miss}}

\newcommand\sigmab{\Delta_b}
\evensidemargin=\oddsidemargin
\def\dsp{\displaystyle}

\def\g {\gamma}

\def\bar {\overline}

\def\bea {\begin{eqnarray}}
\def\eea {\end{eqnarray}}

\def\rdrdst {R(D^{(*)})}

\def\rk {R_K}
\def\rkst {R_{K^*}}
\def\rkrkst {R_{K^{(*)}}}
\def\rJpsi{R_{J/\psi}}

\def\g{\gamma}

\def\beq{\begin{equation}}
\def\eeq{\end{equation}}
\def\barr{\begin{array}}
\def\earr{\end{array}}
\def\dis{\displaystyle}

\def\gev{\ensuremath{\mathrm{Ge\kern -0.1em V}}}
\def\tev{\ensuremath{\mathrm{Te\kern -0.1em V}}}

\def\Babar{{\mbox{\slshape B\kern-0.1em{\smaller A}\kern-0.1em B\kern-0.1em{\smaller A\kern-0.2em R}}}}
\definecolor{darkgreen}{cmyk}{1,0,1,0.4}

\definecolor{pink}{cmyk}{0.4,1,0.3,0}
\def\com2#1{\textcolor{pink}{\large(\it{#1})}}
\begin{document}
\begin{center}
{\Large \bf Search for opposite sign muon-tau pair and 
a b-jet at LHC in the context of flavor anomalies}\\
\vspace*{0.5cm} 
{\sf $^{a}$Debajyoti Choudhury\footnote{debajyoti.choudhury@gmail.com},
          ~$^{a,b,c}$Nilanjana Kumar\footnote{nilanjana.kumar@gmail.com}, 
          ~$^{d}$Anirban Kundu \footnote{akphy@caluniv.ac.in}} \\
\vspace{10pt} {\small } 
{\em $^{a)}$Department of Physics and Astrophysics, University of Delhi, Delhi 110007, India\\
$^{b)}$Saha Institute of Nuclear Physics, HBNI, 1/AF Bidhan Nagar, Kolkata 700064, India\\ 
$^{c)}$The Institute of Mathematical Sciences, HBNI, Taramani, Chennai 600113, India \\
$^{d)}$Department of Physics, University of Calcutta, 92 A.P.C Road, Kolkata 700009, India}
\normalsize
\end{center}
\bigskip
%===========================================================================================================================================
\begin{abstract}
Extant anomalies in several semileptonic $B$-meson decays argue
for physics beyond the Standard Model. Measurements of both
neutral-current decays (such as $R_K$, $R_{K^*}$ and $B_s\rightarrow
\phi\mu\mu$) as well as charged-current ones---$R(D)$ and
$R(D^*)$---provide strong hints for the violation of lepton flavor
universality. Recent studies
(Refs.\ \cite{Choudhury:2017qyt, Choudhury:2017ijp}) have shown that a
class of effective field theory (EFT) models may explain such
anomalies in terms of only a few parameters which can be determined
phenomenologically. In this literature, we examine such 
resolutions in the context of the requisite 
$(\bar{s}$\ $b$)($\bar\tau$\ $\tau$) operator, and look for its signals
at the 13 TeV LHC, with a final state of one $b$-jet, and an
oppositely charged $\mu$-$\tau$ pair, with the muon coming from the
decay of one of the $\tau$ leptons. We obtain discovery and exclusion 
limits on the model parameters as a function of luminosity at the 13 TeV LHC.
\end{abstract}
\bigskip
%========================================================================================================================================
\section{Introduction}
%========================================================================================================================================

Evidence from a multitude of experiments, such as
\Babar~\cite{Lees:2013uzd}, Belle~\cite{Huschle:2015rga,Abdesselam:2016cgx,
Hirose:2016wfn, Sato:2016svk,Hirose:2017dxl,Abdesselam:2019wac}
and, most recently, LHCb~\cite{Aaij:2015yra,Aaij:2017uff,Aaij:2017deq,
Aaij:2014ora,Aaij:2017vbb,CERN-EP-2019-043} suggest the presence of 
effects that violate lepton flavor universality, a cardinal
principle within the Standard Model (SM). Related to several mesons
containing the bottom (anti-)quark, these ``anomalies'' appear in both
charged-- and neutral-current (NC) decays. For example, consider the
ratios ~\cite{Amhis:2016xyh}
%============================
\beq
\label{eq:RD}
\rdrdst \equiv \frac{ {\rm BR}(B\to D^{(*)}\tau\nu)}{ {\rm BR}(B\to D^{(*)}\ell \nu)}\,,
\eeq
with $\ell=e$ or $\mu$, and, similarly, 
\beq
\label{eq:defrjpsi}
\rJpsi \equiv \frac{ {\rm BR}(B_c\to J/\psi\, \tau\nu)}{ {\rm BR}(B_c\to J/\psi\, \mu \nu)}\,.
\eeq
%===========================
While the SM estimates for the individual decays are already quite
robust, the advantage of considering such ratios is that much of the
remaining uncertainties, residing in the evaluation of the form
factors, cancel out. Thus, any observed anomaly in such areas
would be very intriguing, and we begin by recalling the experimental status.

The \Babar~\cite{Lees:2013uzd} measurements of $R(D)$ and $R(D^{*})$,
when taken together, exceed SM expectations by more than $3 \sigma$.
Though the Belle measurements \cite{Huschle:2015rga} lie a little below
the \Babar~measurements and are consistent with both the latter and
the SM expectations, their result on $R(D^{*})$
\cite{Abdesselam:2016cgx}, with the $\tau$ decaying semileptonically,
agrees with the SM expectations only at $1.6 \sigma$ level.
Similarly, the LHCb measurement \cite{Aaij:2015yra} lies $2.1 \sigma$
above the SM predictions. Taking into account all the measurements
and their correlations, the disagreement between the data and SM is at
nearly $3.8\sigma$ \cite{Amhis:2014hma,hflav}. 

The four-fermi effective interaction responsible for $B\to K^{(*)} \ell \nu$,
namely $b \to c \ell \nu$ is also responsible for driving $B_c \to J/\psi \ell \nu$ and 
analyzing $3\, {\rm fb}^{-1}$ data. The LHCb Collaboration found
%=========================================
\beq
\label{data:RJpsi}
\rJpsi= 
\left\{\hskip-5pt
\barr{lcl}
\dsp 0.71 \pm 0.17\pm 0.18 &\hskip-5pt & ({\rm exp.}),
\\[2ex]
\dsp0.283 \pm 0.048 &\hskip-5pt & ({\rm SM}), 
\earr
\right.
\eeq
where the SM prediction~\cite{Ivanov:2005fd,Dutta:2017xmj,Watanabe:2017mip}
includes uncertainties accrued from the $B_c\to J/\psi$ form
factors and is, thus, quite robust. While the level of the discrepancy 
is only at the $2\sigma$ level\footnote{Given the smaller 
production cross section, the large uncertainty is understandable. This 
is expected to improve a lot once more data is analyzed.}
it is interesting to note that it points in the same
direction as the others. 

An opposite effect is seen for the neutral current transitions,
namely $b \to s \ell^+ \ell^-$. Once again, ratios of such decays 
constitute robust variables leading us to consider
\beq
\label{eq:RK}
\rkrkst \equiv \frac{ {\rm BR}(B\to K^{(*)} \mu^+\mu^-)}{ {\rm BR}(B\to K^{(*)}e^+ e^-)}\,.
\eeq
While the SM predictions for both $\rk$ and $\rkst^\text{\,central}$ are
almost indistinguishable from unity
~\cite{Hiller:2003js,Bobeth:2007dw,Bordone:2016gaq,Capdevila:2016ivx,Serra:2016ivr}, 
that for $\rkst^\text{\,low}$ is $\sim$ 0.9 (mainly due to the 
non-negligible $m_\mu$). The calculations are very precise with
only minuscule uncertainties. The earlier result on $\rk$~\cite{Aaij:2014ora,Aaij:2017vbb} 
has recently been superseded by the LHCb Collaboration~\cite{CERN-EP-2019-043}:
%=========================================
\beq
\label{data:RK}
\barr{rclcl}
\rk &=& \dis 0.846^{+0.060+0.016}_{-0.054-0.014} & \qquad &\dis q^2 \in [1.1:6]\, {\rm GeV}^2\,,\\[2ex]
\earr
\eeq
%==========================================
For $\rkst$, while the earlier LHCb data~\cite{Aaij:2014ora,Aaij:2017vbb}, viz.
%==========================================
\beq
\barr{rclcl}
\rkst^\text{\,low} &=&\dis {0.66}^{+0.11}_{-0.07} \pm 0.03 & \qquad &\dis  q^2 \in [0.045:1.1]\, {\rm GeV}^2\,,
\\[2ex]
\rkst^\text{\,central} &=& \dis 0.69 ^{+0.11}_{-0.07} \pm 0.05  & \qquad &
        q^2 \in [1.1:6]\, {\rm GeV}^2\,,
\earr
\eeq
%=========================================
were significantly away from the SM predictions, the recent Belle
results~\cite{Abdesselam:2019wac}, taking average over $K^{*0}$ and $K^{*+}$ modes,
are more compatible with the SM:
%=========================================
\beq
\label{data:RKST}
\barr{rclcl}
\rkst &=& \dis 0.94 ^{+0.17}_{-0.14}\pm0.08 & \qquad &\dis q^2 \in [0.045,\rm]\, {\rm GeV}^2\,.
%\rkst^\text{\,low} &=&\dis {0.66}^{+0.11}_{-0.07} \pm 0.03 & \qquad &\dis  q^2 \in [0.045:1.1]\, {\rm GeV}^2\,,\\[2ex]
%\rkst^\text{\,central} &=& \dis 0.69 ^{+0.11}_{-0.07} \pm 0.05  & \qquad &
%        q^2 \in [1.1:6]\, {\rm GeV}^2\,.
\earr
\eeq
%=========================================
While this can be construed as the $\rkst$ average 
moving closer to the 
SM (note, though, the larger errors in the Belle results), an
anomaly is still hinted at, with the magnitude of the deviation
being somewhat less than that in the $\rk$ data.

A corroborating deviation is seen in $B_s\to \phi\mu\mu$~\cite{Aaij:2015esa,Altmannshofer:2014rta,Straub:2015ica},
namely,
%=========================================
\beq
\label{data:phimumu}
\frac{d~}{dq^2}{\rm BR}(B_s\to\phi\mu\mu){\Big|}_{q^2\in[1:6]\,{\rm GeV}^2}
= \left\{ 
    \barr{lcl}
    \dis \left(2.58^{+0.33}_{-0.31}\pm 0.08\pm 0.19\right) \times 10^{-8}~{\rm GeV}^{-2} & \quad & ({\rm exp.}) \\[2ex]
    \dis \left(4.81 \pm 0.56\right) \times 10^{-8}~{\rm GeV}^{-2} & & ({\rm SM})\,. 
  \earr
  \right.
\eeq
%==========================================
where $q^2 = m^2_{\mu\mu}$. This suggests that the discrepancies in
$\rk$ and $\rkst$ have been caused by a depletion of the $b \to s
\mu^+ \mu^-$ channel, rather than an enhancement in $b \to s e^+ e^-$.
Such a conclusion is lent further weight by the long-standing $P'_5$
anomaly~\cite{LHCb:2015dla} in the angular distribution of $B\to
K^*\mu\mu$, with a more than $3\sigma$ mismatch between the data and
SM prediction\footnote{ Recently, however, it has been 
argued in Ref~\cite{Datta:2019zca} that these discrepancies 
may have their origin, instead, in some new physics in 
$b\to s e^+e^-$.}.

Faced with all these anomalies, two approaches are possible. The
first would be to construct an elaborate ultraviolet-complete
theory. Examples are offered by $Z'$-models~\cite{Langacker:2008yv}
with flavor violating couplings with the quarks and the
leptons~\cite{Falkowski:2015zwa,Buras:2013qja,Kamenik:2017tnu,Carmona:2015ena,
Chala:2018igk,Chala:2019vzu}
on the one hand, and, on the other, the exchange of
leptoquarks~\cite{Becirevic:2016yqi,Fajfer:2015ycq,Blanke:2018sro} or, equivalently,
sfermions in R-parity violating supersymmetric
models~\cite{Barbier:2004ez,Farrar:1978xj,Altmannshofer:2017poe}. The alternative is to take
recourse to an effective field theory (EFT) description wherein only a
set of Wilson coefficients are altered from their SM
values~\cite{Alonso:2016oyd,Bardhan:2017xcc,Bobeth:2001sq,
Altmannshofer:2008dz,Crivellin:2015lwa,Das:2016vkr,Bhattacharya:2016mcc}.
In either case, one would, naively, expect that a sufficiently large
set of unknown parameters (and/or fields) would need to invoked so as
to enable the simultaneous explanation of all the anomalies while
maintaining the rest of the well-tested SM phenomenology. However, if
the nature of the UV-theory (operative at a scale higher than the
electroweak scale), the integrating out of whose heavy degrees of
freedom is supposed to have given us the EFT, is entirely ignored,
then a phenomenologically motivated EFT with only a small number of
parameters need to be considered. It has been
shown~\cite{Choudhury:2017qyt, Choudhury:2017ijp,Bhattacharya:2019eji}
that such a minimal set of new physics (NP) operators, accompanied by a single
lepton mixing angle, can indeed explain almost all the observables
adequately. More interestingly, the natural scale for such an
explanation is seen to be a few TeVs, opening the interesting
possibility of signatures at the LHC and/or future colliders.

In the present case, instead of attempting a generic study, we
consider a particular signature, at the LHC, prompted by the scenarios
discussed in Refs.~\cite{Choudhury:2017qyt, Choudhury:2017ijp}. Some
such studies have been attempted in the past, but in entirely
different contexts, both at the simulation level
~\cite{Afik:2018nlr,Allanach:2018odd,Cerri:2018ypt,
 Greljo:2017vvb,Arbey:2018ics,PhysRevD.94.075006,Huang:2018nnq,Faroughy:2016osc},
as well as by the ATLAS and CMS collaborations who have searched for
flavor violating signatures with dilepton final
states~\cite{Aad:2012ypy,Aaboud:2016hmk,Aaboud:2018jff,
 ATLAS:2017wce,Aad:2015gha,Sirunyan:2018jdk}. We focus on a model
which can explain, simultaneously, both the CC and the NC anomalies in
semileptonic $B$ decays, for example, one with an enhanced
$(\bar{s}$\ $b$)($\bar\tau$ $\tau$) operator. We have studied the
tell-tale signatures which include an opposite signed $\mu$-$\tau$
pair and a $b$-jet, induced by this operator when one tau decays to a
muon. One of the novelties of this channel is that it does not suffer
from a very large background unlike opposite sign same flavor
lepton-pair signatures. Apart from effecting a full simulation, 
we also validate our background estimation with 
Ref.\cite{Sirunyan:2018jdk}. This, as well as our choice of a robust 
set of observables renders our methodology applicable 
to a very wide class of NP scenarios.

The rest of the paper is constituted as follows. In 
the next section, we briefly discuss the EFT framework. In
Section 3, we study the collider signatures of the particular channel
with detail analysis of the signal and background at 13 TeV LHC. Then,
in Section 4, we discuss the discovery and exclusion perspectives of
this particular channel. Lastly, we conclude by predicting some
future possibilities in Section 5.
%===========================================================================================================================================
\section{Effective Theory Model}
%===========================================================================================================================================

Considering all new physics (NP) effects to be parametrized by
$SU(3)_c \otimes SU(2)_L \otimes U(1)_Y$ invariant four-fermi
operators, one needs at least two such
structures~\cite{Choudhury:2017ijp} so as to both explain the
anomalies and be consistent other low-energy observables. While
Refs.~\cite{Choudhury:2017qyt, Choudhury:2017ijp} did consider several
possibilities, they identified certain combinations as favored
scenarios. Subsequently, Ref.~\cite{Bhattacharya:2019eji} reexamined
the data, taking into account all correlations and, apart from
establishing these scenarios (modulo certain alterations in the
allowed parameter space), found that even the combinations dismissed
in Ref.~\cite{Choudhury:2017qyt} can be accommodated. Rather than
examine each such scenario, we consider a particular representative
case, termed ``Model IV'' in Ref.~\cite{Choudhury:2017qyt}. Analyses
for the other scenarios can also be effected analogously. The
Hamiltonian for the new physics can be expressed in terms of two
operators involving left handed doublets $Q_{2L}$, $Q_{3L}$ and
$L_{3L}$ and right handed singlet $\tau_{R}$ as,
%%%%
\beq
\barr{rcl}
{\cal H}^{\rm NP} & = & \dis 
   \sqrt{3} \, A_1 \, \left[
- (\bar Q_{2L} \g^\mu Q_{3L})_3 \, (\bar L_{3L} \g^\mu L_{3L})_3
+ \frac{1}{2} \, (\bar Q_{2L} \g^\mu L_{3L})_3 \, (\bar L_{3L} \g^\mu Q_{3L})_3
     \right]
\\[2ex]
&+ & \dis
\sqrt{2} \, A_5 \,  (\bar Q_{2L} \g^\mu Q_{3L})_1 \, (\bar \tau_R \g^\mu \tau_R) \ + h. c.,
\earr
    \label{eq:H_NP1}
\eeq
%=============
where $A_{1,5}$ are unknown coefficients of mass dimension $-2$ and to be
determined phenomenologically. For the sake of simplicity, we assume
these to be real. The subscripts `3' and `1' represent the
$SU(2)_L$ triplet and singlet currents respectively.
These can be expressed in terms of component fields as
%%%%
\beq
\barr{rcl}
{\cal H}^{\rm NP}\ 
 & = & \dis \frac{3 \, A_1}{4} \, (c, b) \, (\tau, \nu_\tau)
+ \frac{3 \, A_1 }{4} (s, b) (\tau, \tau) + A_5 \, (s, b) \, \{\tau, \tau\} 
\\[2ex]
& + & \dis \frac{3 \, A_1}{4} \, (s, t) \, (\nu_{\tau}, \tau) 
   + A_5 (c, t) \{\tau, \tau\}
   + \frac{3 \, A_1 }{4}(c, t) \, (\nu_\tau, \nu_\tau)+ h.c.,
\earr
    \label{eq:H_NP2}
\eeq
%==============
where, following the notation introduced in~\cite{Choudhury:2017ijp}, 
we denote
%%%
\beq
(x,y) \equiv \bar x_L \g^\mu y_L ; \ \ \ \{x,y\} \equiv \bar x_R \g^\mu y_R \ \ \ \  \forall\ \  x,y\,.
\eeq
%==============
Note that ${\cal H}^{\rm NP}$ is expressed in terms of weak
eigenstates involving the second and the third generation quark
fields ($Q_{2L}$ and $Q_{3L}$), but only the third generation
leptons ($L_{3L}$). While the quark fields would be affected by the
usual CKM mixing, in the leptonic sector, the weak eigenstates can be
related to the mass eigenstates through a further field
rotation~\cite{Choudhury:2017qyt,Choudhury:2017ijp}.
This, of course, would induce direct lepton flavor violation.
The magnitude of this mixing, as deduced
phenomenologically~\cite{Choudhury:2017qyt,Choudhury:2017ijp}, is,
however small, and, was perfectly consistent with
$Br (B^+ \to K^+ \mu^\pm\tau^\mp) < 4.8\times 10^{-5}$ (at 90\% C.L.)~\cite{Tanabashi:2018oca}.
Indeed, it also easily satisfies the
recently quoted 95\% C.L. upper bound of $Br (B_s \to \tau^\pm
\mu^\mp) < 4.2 \times 10^{-5}$~\cite{Aaij:2019okb}. This very
smallness of the mixing allows us to 
neglect it altogether and concentrate on the operator $(s, b) (\tau,
\tau)$ and $(s, b) \, \{\tau, \tau\}$ alone\footnote{It might be
argued that, on inclusion of further quantum corrections, this
operator can adversely affect the $B_s$-$\bar{B_s}$ mixing. This
issue has been adequately addressed in Ref.~\cite{Choudhury:2017ijp}.}.

If an ultraviolet-complete origin of Eq.(\ref{eq:H_NP2}) is desired,
an individual term could be parametrized as,
\[
\label{eq:mix3}
X(a,b)(c,d) = {\frac{\lambda_1^*\lambda_2}{2 \, M^2}}\, (a,b)(c,d)\,,
\]
where $M$ is the mass of the integrated-out field, and $\lambda_i$ are
some dimensionless couplings, bounded from perturbativity by
$\lambda^2/(4\pi)^2 \leq {\cal O}(1)$. The mediator, for example, might be a
leptoquark, or a $Z'$ with flavor-changing couplings.
This inequality, alongwith the requirement of reproducing the requisite 
$A_i$ would determine the ranges allowed to $\lambda_i$ and $M$.
We, however, eschew any assumption as to the UV-completion, 
resolutely choosing to be agnostic as to the origin of the $A_i$.
%============================================================
\begin{table}[tpb]
  \begin{center}
    \caption{\label{tab:1}\em Benchmark regions to study $(s, b) (\tau, \tau)$ and $(s, b) \, \{\tau, \tau\}$ 
operators in ($\mu^\pm \tau^\mp$) pair + $b$-jet final state, based on Ref.~\cite{Bhattacharya:2019eji}.}
\begin{tabular}{|c|c|c|c|c|c|}
\hline
\hline
Observables & Set X & Set Y & Set A & Set B & Set C \\
\hline
Br$(B_s\rightarrow \tau \tau)< 6.8\times10^{-3}$ & \cmark &\cmark &  \cmark   &\cmark & \cmark\\
\hline
$3\sigma$ contour around  &&&&&\\

$A_1 \approx -3.8$, $A_5 \approx -2.3$           & \cmark  & \xmark   &  \xmark  & \xmark    & \xmark\\
\hline
\hline
\end{tabular}
\end{center}
\end{table}
%========================================================================

While Refs.\cite{Choudhury:2017ijp, Bhattacharya:2019eji} do zero in
on ``best-fit'' points in the parameter space\footnote{The best fit
values were obtained under the assumption of flavor mixing, but our
analysis is independent of the mixing angle. Note, too, that the
best fit values of Ref.~\cite{Choudhury:2017ijp} and
Ref.~\cite{Bhattacharya:2019eji} are very similar.}, note that the
exact location of the same is dependent on the accumulation of more
data and, indeed, even the very recent measurements would change it to
an extent. Consequently, we investigate the LHC signal for a variety
of points, though laying special emphasis to the best fit point of
Ref.~\cite{Bhattacharya:2019eji}, namely, $A_1 \approx -3.8$ and $A_5
\approx -2.3$ and consider several benchmark regions as listed in
Table~\ref{tab:1}.

We consider two factors while defining the
regions: one is the 3$\sigma$ contour around the best fit point as
obtained in Ref.~\cite{Bhattacharya:2019eji} and another is the
current 95\% C.L. limit on Br$(B_s\rightarrow \tau \tau)<
6.8\times10^{-3}$ ~\cite{Aaij:2017xqt}. All the benchmark points
satisfy the latter limit. Set X includes points inside the $3\sigma$
contour keeping $|A_1|$ fixed at 3.8. Set Y represents points just
outside the $3\sigma$ contour around the best fit. Set A, Set B and
Set C constitute regions with smaller values of $|A_1|$ and $|A_5|$
and, hence, represent more {\em conservative} choices, both in the
context of low-energy observables as well as LHC signals. The exact
locations of the points are detailed in Table~\ref{tab:2} in the next section, 
where we study the corresponding collider signals originating 
from ($s,b$)($\tau$,$\tau$) and ($s, b$)$\{\tau, \tau\}$.
%====================================================
\section{Collider study of ($\mu^\pm \tau^\mp$) pair and a $b$-jet at $\sqrt{\hat s}=$13 TeV}
\begin{figure}[!h]
\begin{center}
\begin{minipage}[]{0.4\linewidth}
\includegraphics[width=5.5cm,height=4.4cm]{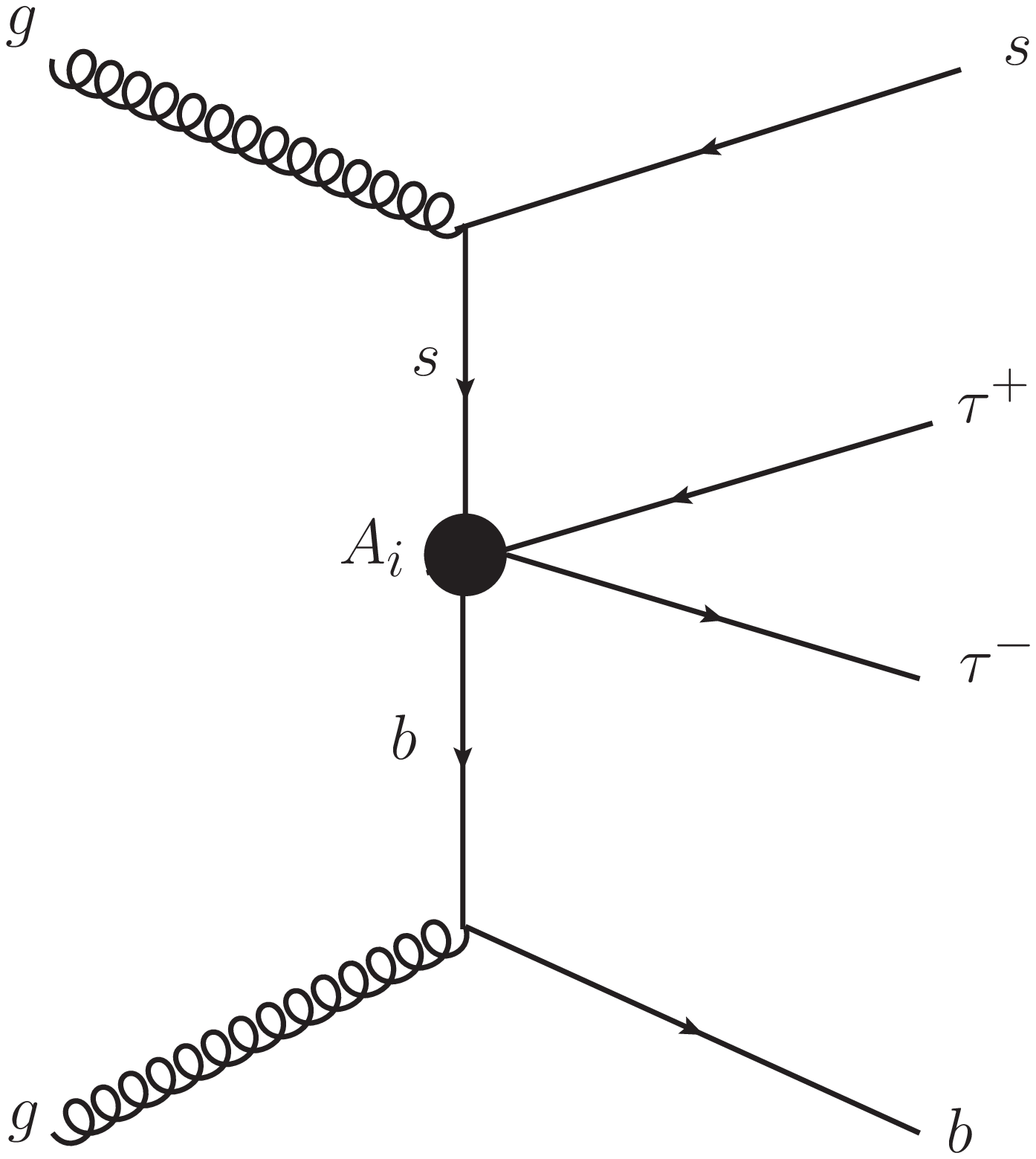}
\end{minipage}
\begin{minipage}[]{0.4\linewidth}
\includegraphics[width=5.8cm,height=2.8cm]{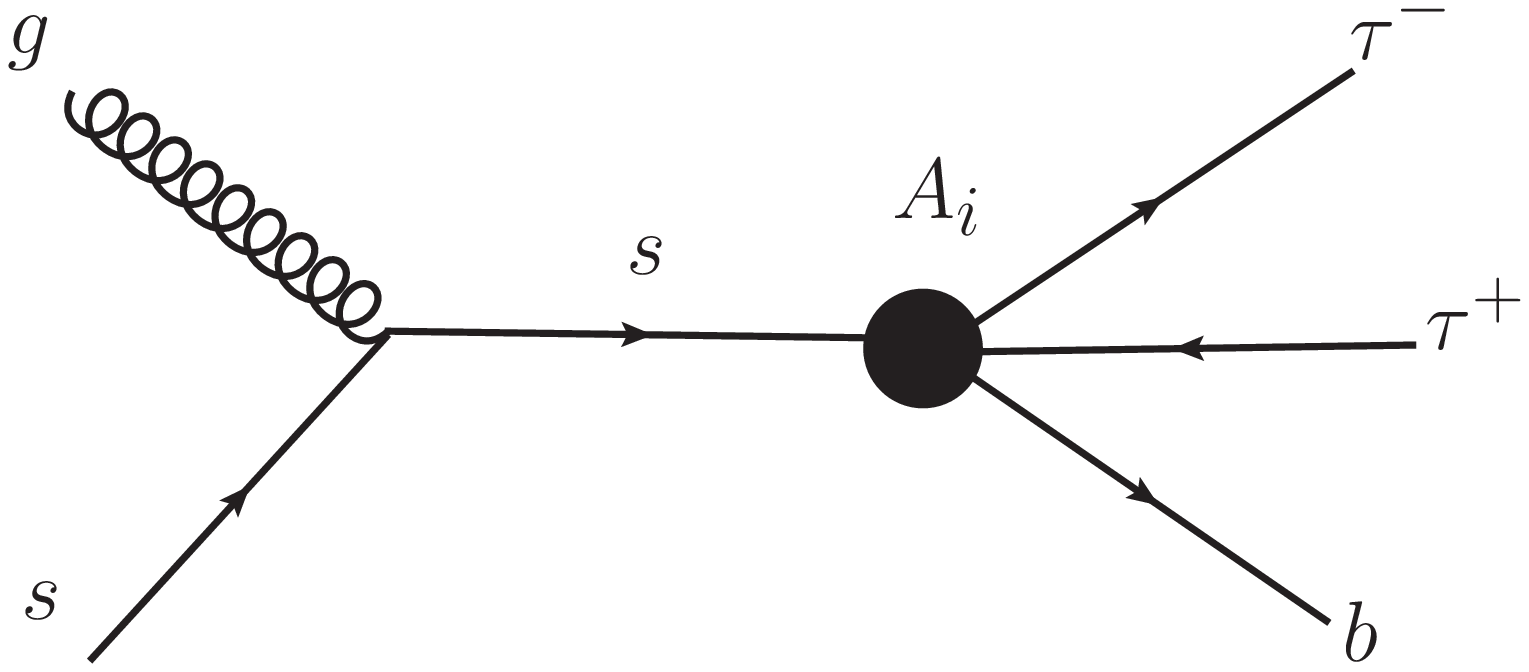}
\end{minipage}
\end{center}
\caption{\label{fig:1} (L) {\it Feynman diagram for the production of 
two $\tau$'s in association with one $b$-jet in $g$-$g$ fusion 
$[g \ g\rightarrow \tau^\pm \tau^\mp b (s)]$,} (R) 
{\it and in $g$-$s$ fusion $[g \ s\rightarrow \tau^\pm \tau^\mp b]$. In both 
figures, the solid blob represents the four-point vertex with 
effective couplings $A_i$.}}
\end{figure}

%====================================================
The signature of our interest, namely, a $\mu^\pm \tau^\mp$ pair
accompanied by a $b$-jet, originates from the operators
($s,b$)($\tau$,$\tau$) and ($s, b$)$\{\tau, \tau\}$ in
Eq.~(\ref{eq:H_NP2}). The requirement of one additional $b$-jet with
opposite sign lepton pair reduces the SM background significantly. The
muon, for the signal events, emanates from the decay of a
$\tau$. While a direct production of a $\mu^\pm \tau^\mp$ pair is
possible if the aforementioned lepton-mixing is nonzero, the very
smallness of the corresponding angle renders this channel to a very
subdominant role\footnote {If one considers flavor mixing, there can
be three signatures: ($\mu^\pm \mu^\mp$) pair + $b$-jet, ($\mu^\pm
\tau^\mp$) pair + $b$-jet and ($\tau^\pm \tau^\mp$) pair + $b$-jet,
each with effective coupling as a function of flavor mixing
angle. The signature ($\mu^\pm \mu^\mp$) pair + $b$-jet has been
studied in detail recently~\cite{Afik:2018nlr}.}. Consequently, we
neglect the mixing altogether, even though it is relevant to explain
the anomalies. The final states $\tau^\pm \tau^\mp b \ s$ and
$\tau^\pm \tau^\mp b$ can be produced from $g$-$g$ fusion and $g$-$s$
fusion respectively in $p$-$p$ collision, as shown in
Fig.~\ref{fig:1}, and both the processes are considered in our
analysis.
%===============================================
%================================================
In the next section we determine the sensitivity of the LHC
in the recently concluded run as well as in the forthcoming one for
the model parameters, $A_i$. Towards this, we analyze
the signal for each of the benchmark regions as listed in Table~\ref{tab:1}.

Given the preferred size of the four-fermi couplings $A_i$, the
channel ($\tau^\pm \tau^\mp$) pair + $b$-jet has a large production
cross section, even if we demand that one of the $\tau$'s decays into
a $\mu$ (with a branching fraction of 0.174). On the other hand,
owing to the smallness of the lepton mixing angle, the cross section
for direct production, from $g$-$g$ or $g$-$s$ fusion, of $\mu^\pm
\tau^\mp$ + $b$-jet is very small. In the four-fermi limit, the
production cross-section would depend only on the couplings $A_i$ 
with the subprocess cross-section 
scaling simply as $\hat s \, A_i^2$ 
where $\sqrt{\hat s}$ is the subprocess center-of-mass
energy. This, of course, is moderated by the $\sqrt{\hat s}$-dependent
parton flux. However, if an ultraviolet-complete theory is considered
instead, the dependence on $\hat s$ and the mediator mass scale
$M$ is more complicated, and depends on the
precise nature of the completion (for example, a $Z'$-like theory
would admit the possibility of a resonance, while a leptoquark-like
theory would only have $t$-channel propagators). In addition, the
phase space distributions would differ as well. However, for a
mass-scale $M$ that is larger than a few TeVs, these differences
quickly subside primarily on account of the relevant parton-fluxes
falling quickly with $\sqrt{\hat s}$. Not only does this result in a
suppression of the fraction of events that could potentially be
sensitive to a possible resonance, but any such resonance would also
be relative wide one, given the preferred values for the $A_i$.

The new physics here is simple enough to permit an analytic
calculation, which, when followed by a simplistic simulation, yields
rather robust results. Nonetheless, we also implement the effective
theory model in Feynrules~\cite{Alloul:2013bka,Christensen:2008py} and
generate signal events, at the leading order, uniformly throughout the
parameter space with MadGraph5\_aMC@NLO (v2.2.1)~\cite{Alwall:2011uj}
interfaced with PYTHIA~\cite{Sjostrand:2006za}. For this, we use the
NNPDF23LO1~\cite{Ball:2012cx} parton distributions with the 5-flavor
scheme. We kept the factorization scale fixed at $m_T^2$ after $k_T$
-clustering of the event. On varying the factorization scale between
$m_T^2/4$ and $4m_T^2$ and, similarly, scanning over different parton
distribution sets, the uncertainty in the LO signal cross section is
found to be less than 16\% and 10\% respectively, for the entire range
of the parameter space. As for the NLO corrections to the SM, the
calculation thereof for a scattering process with a multibody final
state such as ours, and, especially in an effective theory, is a very
arduous task and beyond the scope of the present work. An intelligent
estimate can be made nonetheless, by realizing that the effective
Hamiltonian of Eq:\ref{eq:H_NP2} is most easily obtained starting from
a theory with a flavour-changing $Z'$ or one with scalar
leptoquarks. The $K$-factor for the former is about
1.3~\cite{Fuks:2007gk, Fuks:2017vtl} while that for the latter is in
the range 1.3--1.4~\cite{Dorsner:2018ynv}. It should be appreciated,
though, that the relevant couplings considered in the said references
are not exactly what are needed for the present case. Nevertheless, it
stands to reason that the exact $K$-factor should not be wildly
different from those quoted above. In other words, the higher-order
corrections are expected to increase the signal cross sections. We,
however, adopt a {\em conservative} standpoint in choosing not to
include this enhancement.

The signal comprises two processes, namely, $g \ s\rightarrow \tau^\pm
\tau^\mp \ b$ and $g \ g\rightarrow \tau^\pm \tau^\mp \ b \ s$ and
only those events are selected wherein one tau decays leptonically to
muon, resulting in a ($\mu^\pm \tau^\mp$)+ $b$-jet final state. The
events are passed through DELPHES~3~\cite{deFavereau:2013fsa}, in
order to incorporate detector effects and apply reconstruction
algorithms. Jets are reconstructed using the anti-$k_T$ algorithm in
FastJet~\cite{Cacciari:2011ma}. For muon isolation, we have required
$\Delta R \geq 0.4$ and $p_T > 1$ GeV. For calorimetric (tracking)
isolation, we require the corresponding momentum parameter to be 0.14
(0.15) times the $p_T$. This ensures that the muons are well isolated
from other objects. Tau leptons are reconstructed through their
hadronic decays, and we demand that $\Delta R \geq 0.4$ and $p_T > 10$
GeV for the reconstruction. In DELPHES, the tau-tagging efficiency is
considered to be 0.6 and tau misidentification (from gluons and
quarks) probability is 0.001. In this analysis, jets are required to
have $p_T$ greater than 30 GeV and $\eta(b)< 4.7$, and must be
separated from the selected leptons by $\Delta R\geq 0.5$. For tagging
the $b$-jets, we used a $b$-tagging module inside DELPHES with 70\%
working efficiency. The probability of misstagging a charm as $b$-jet
is 10$\%$ while for the other quarks and gluons, it is 0.1$\%$ or
less.

The backgrounds for this channel can be classified into two
categories. The irreducible backgrounds arise
mainly from $t \overline{t}$, single top ($Wt$), and diboson
($W^+W^-$, $WZ$ and $ZZ$) production, whereas the $t
\overline{t}W$, $t \overline{t}Z$ contributions are very small. 
The major contributions to the reducible background
arise from $W$+jets, $Z/\gamma$+jets and other QCD multi-jet
processes, where jets may be misidentified as
leptons. The probability
of a jet to be misidentified as a lepton is taken as a module inside
DELPHES ~\cite{Khachatryan:2014sta}, as a function of $p_T$ and $\eta$ of 
the jet. All
background events are generated using MadGraph and the cross-sections
are taken upto NLO and upto NNLO in some cases (see Ref.~\cite
{ATL-PHYS-PUB-2019-010} and references within). Most
of the backgrounds will be reduced by the requirement of one $b$-jet
and strong isolation selection among the opposite sign leptons. 

As we shall see later, the cuts we propose are very effective in
suppressing the large background (from a host of SM processes) and,
thereby, in increasing the signal-to-noise ratio. To eliminate large
errors in background modelling, we effect a comparison with an actual
experimental study. In particular, the CMS collaboration has performed
a search~\cite{Sirunyan:2018jdk} for a singly produced
third-generation scalar leptoquark decaying to a tau lepton and bottom
quark in proton-proton collisions at 13 TeV which includes the final
state ($\mu^\pm \tau^\mp$) pair + $b$-jet. The selections we imposed
can be summarized as follows: Exactly one each of a $\tau$ and a $\mu$
with these being oppositely charged and with a single $b$-jet,
satisfying,
\begin{center}
\underline {Selection {\bf S0}}:\\
$p_T(\mu)> 50$ GeV, $p_T(\tau)> 50$ GeV, $p_T(b)> 50$ GeV,\\
$\eta(\mu)< 2.4$, $\eta(\tau)< 2.5$, $\eta(b)< 2.4$,\\
$\Delta R(\mu,\tau)> 0.5$, $\Delta R(\mu,b) > 0.5$ .
\end{center}
%=======================
\begin{figure}[!h]
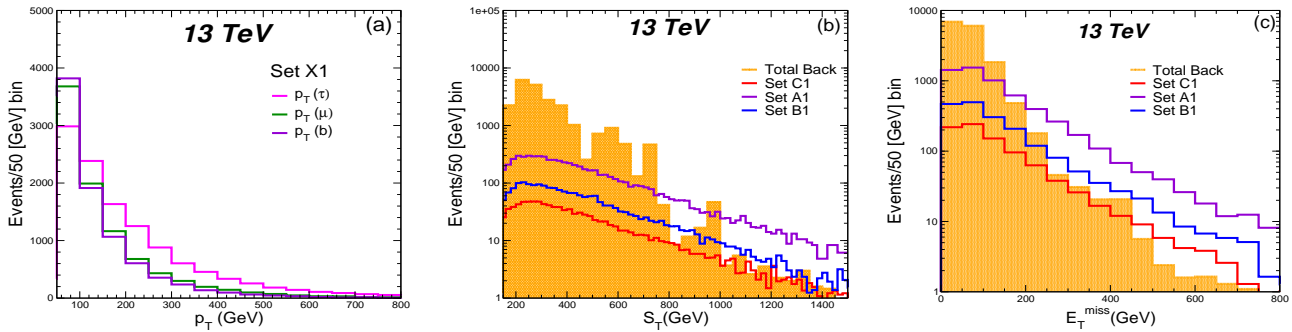

\begin{center}
\includegraphics[width=5.3cm,height=4.3cm]{pt.eps}
\hskip 10pt
\includegraphics[width=5.3cm,height=4.3cm]{st.eps}
\hskip 10pt
\includegraphics[width=5.3cm,height=4.3cm]{met.eps}
\end{center}
\caption{\label{fig:2} {\em (a)} $p_T$ distributions of the signal
($\mu^\pm \tau^\mp$) pair + $b$-jet. {\em (b)} $S_T=p_T(\mu)+ p_T(\tau)+ p_T(b)$ distribution 
for the signal and total background {\it {\em (c)}
$\missET$ for the signal and total background. All plots are done after 
after Selection ${\bf S0}$. The signals are normalized for Set X1 ($|A_1|=3.8$ and $|A_5|=2.3$)
C1 ($|A_1|=1.0$ and $|A_5|=0.5$), A1 ($|A_1|=3.0$ and $|A_5|=1.5$), B1 ($|A_1|=2.0$ and $|A_5|=1.0$). 
Events are weighted at 120 fb$^{-1}$.}}
\end{figure}
%=======================

\begin{figure}[!h]
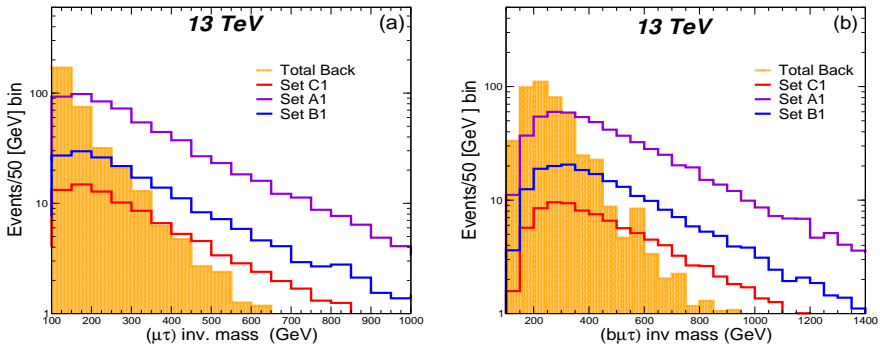

\begin{center}
\includegraphics[width=5.5cm,height=4.5cm]{mutau.eps}
\hskip 10pt
\includegraphics[width=5.5cm,height=4.5cm]{dist.eps}
\end{center}
\caption{\label{fig:3} {\it The behaviour of the (a) two body and (b) three body invariant
mass of distribution of the signal ($\mu^\pm \tau^\mp$) pair + $b$-jet for C1, A1 and B1 
(defined in Table~\ref{tab:2}) and the total background is shown after the
${\bf S1}$ cut. Events are weighted at 120 fb$^{-1}$.}}
\end{figure}

Selection ${\bf S0}$ constitutes essentially the basic cuts on
different variables. The cuts on $\Delta R$ is placed to ensure that
the muon is well isolated from the tau and the b-jet. We have plotted
the respective $p_T$ distributions for the tau, the muon and the
$b$-jet for the signal corresponding to a representative point in the
parameter space in Fig.~\ref{fig:2}(a). As already discussed, the hard
interaction being a four-fermi one, the production cross sections,
typically, grow with the partonic center-of-mass energy (modulo the
suppression due to the effective flux). Furthermore, with the cross
sections slightly favouring large angle scattering over small-angle,
each of the two $\tau$'s as well as the $b$-jet tend to have a
sufficiently large $p_T$. While the $\tau^\pm$ have essentially
identical distributions, the $b$ has a softer component, that arises
from the $gg$-initiated process. The $\mu$, while having considerable
$p_T$, is softer than the $\tau$, being only a descendant of the
second $\tau$. Given this, it is profitable to impose stronger cut in
terms of the variable $S_T$, which is the scalar sum of the $p_T$ of
the final state particles. The distribution is shown in
Fig.~\ref{fig:2}(b) for both signal and the background. The signal is
associated with a relatively modest $\missET$, as seen from
Fig.~\ref{fig:2}(c). As in ~\cite{Sirunyan:2018jdk}, a soft cut on
$M_{\mu\tau}$ in {\bf S1} is essential to reduce the background coming
from the $t \overline{t}$ and single top production. To summarize,
the following set of cuts are used, as in
Ref.~\cite{Sirunyan:2018jdk}.
\begin{center}
\underline{Selection {\bf S1}}:\\ 
$S_T=p_T(\mu)+ p_T(\tau)+ p_T(b) > 500$ GeV,\\
$M_{(\mu,\tau)} > 85$ GeV, \\
\end{center}
The distributions at this stage are depicted in Fig.\ref{fig:3}. It
is worthwhile to point out that our background profile shape agrees
very well with the CMS results~\cite{Sirunyan:2018jdk} while we exceed
them in total count by about 5\%. In other words, our background
determination is very robust.
%==================
%=================================================

At this stage, we propose the following additional selections 
to improve the signal efficiency,
\begin{center}
\underline{Selection {\bf S2}}:\\ 
$\missET$ $ < 230$ GeV,\\
$M_{(b,\mu,\tau)} > 600$ GeV.
\end{center} 
  These particular choice needs some explaining. While the major
  background process ($t \bar t$) has a pair of neutrinos associated
  with the hard process, the signal has none (and the $\missET$ arises
  from cascade decays of the hadron as well as
  mismeasurements). Consequently, a strong upper cut on $\missET$
  would be expected to be useful. In the present case, though, the
  background from the single top and $W$+jets processes are not
  negligible, and these are not reduced overwhelmingly by the
  imposition of such a cut. As a result the loss of background
  resulting from a stronger upper cut on $\missET$ does not compensate
  enough the attendant loss in the signal strength. Hence, we impose
  only a loose upper cut of $\missET < 230$ GeV in {\bf S2}. While the
  entire preceding discussion may seem infructuous in view of
  Fig.~\ref{fig:2}(c), it should be viewed in conjunction with the cut
  on the 3-body invariant mass, $M_{(b,\mu,\tau)}$ which plays a major
  role in preferentially reducing the backgrounds. As
  Fig.~\ref{fig:3}(b) demonstrates, a strong cut on this variable is
  expected to improve the signal-to-noise ratio. Indeed, once this is
  imposed (as in {\bf S2}) it turns out that it is an upper
  restriction on $\missET$ that is more useful rather than a lower
  cut. Moreover these particular set of cuts in {\bf S2} are useful
  when signal cross section is comparatively small for small $A_i$.
  The signal cross sections after the cuts are given in
  Table~\ref{tab:2} for different benchmark points of $A_1$ and $A_5$.
  We have analyzed the signal as a function of $A_1$ and $A_5$, in
  different regions and show the variation in the signal cross
  sections.
%===================
\begin{table}[!h]
  \begin{center}
    \caption{\label{tab:2} {\it Signal cross section of ($\mu^\pm \tau^\mp$) 
pair + $b$-jet after the selections at 13 TeV p-p collision at some benchmark points.
}}
\begin{tabular}{|c|c|c|c|c|c||c|c|c|c|}
\hline
\hline
Set & $|A_1|$& $|A_5|$  &$\sigma$({\bf S1})(fb)& $\sigma$({\bf S2})(fb) & Set & $|A_1|$& $|A_5|$ & $\sigma$({\bf S1})(fb)&$\sigma$({\bf S2})(fb)\\
\hline
Y1&	4.5&	3.0&	44.52&	34.72   & A1&    3.0	&1.5	&10.69 & 8.34\\
Y2&	4.5&	3.8&	70.85&	55.41  & A2&    2.5	&1.5	&6.54	&5.12\\
Y3&	4.0&	3.0&	38.95&	30.17  & A3&    2.0	&1.5	&5.67  &4.39\\
Y4&	4.0&	4.0&	80.32&	62.66  & A4&    1.5	&1.5	&4.24	&3.31 \\
\hline
X1&	3.8&	2.3&	22.78&	17.77  & B1&   2.0     &1.0   &3.36	&2.67\\
X2&	3.8&	3.0&	35.6&	27.84  & B2&   1.5     &1.0   &2.31  &1.8 \\
X3&	3.8&	4.0&	 72.39&  55.83  & B3&   1.0     &1.0   &1.23	&0.98\\
\hline
Y5&	3.5&	2.0&    16.94&	13.21  & C1&   1.0   &0.5    &0.81	&0.63\\
Y6&	3.5&	3.0&    35.65&	27.8 & C2&   0.5   &0.5    &0.22	&0.17\\
Y7&	3.0&	2.3&	18.70&	14.63  & C3&   0.1   &0.5    &0.03	&0.05\\
Y8&	3.0&	3.0&	31.13&  24.09 & C4&   0.05  &0.5    &0.02	&0.01\\
\hline										
\hline
\end{tabular}
\end{center}
\end{table}
%=======================
\begin{table}[tpb]
  \begin{center}
    \caption{\label{tab:3} {\it Background cross sections after several selections at 13 TeV p-p collision.}}
\begin{tabular}{|c|c|c|c|c|}
\hline
\hline
    Background & $\sigma$({\bf S1})(fb)& $\sigma$({\bf S2})(fb) \\
\hline
$t\overline{t}$      & 5.715  &2.78 \\
$Wt$ & 1.132  &0.502 \\
$W+ jets$ &  0.275  &0.241\\ 
$Z/\gamma+jets$ & 0.076   &0.038 \\
$Di$-$boson$    & 0.014   &0.0017\\
$QCD+multi$-$jets$       &  0.0624   &0.016\\

\hline
Total & 7.276	 & 3.6 \\
\hline
\hline
\end{tabular}
\end{center}
\end{table}
%========================

The cut-flow of the backgrounds with the selections is demonstrated in
Table~\ref{tab:3}. Even after the selection ${\bf S2}$, the majority
of the total background comes from $t\overline{t}$ and single top
production. Also note that the signal retains reasonable amount of
events when passed through ${\bf S2}$. The background is notably
smaller for the {\bf S2} selection as compared to that for ${\bf S1}$. Even though the
discovery and exclusion limits can be obtained with both ${\bf S1}$
and ${\bf S2}$, and their behaviour is comparable, in the next section
we show the result when events are selected through ${\bf S2}$.

%==========================================================================================================================================
\section{Results}
%==========================================================================================================================================
\begin{figure}
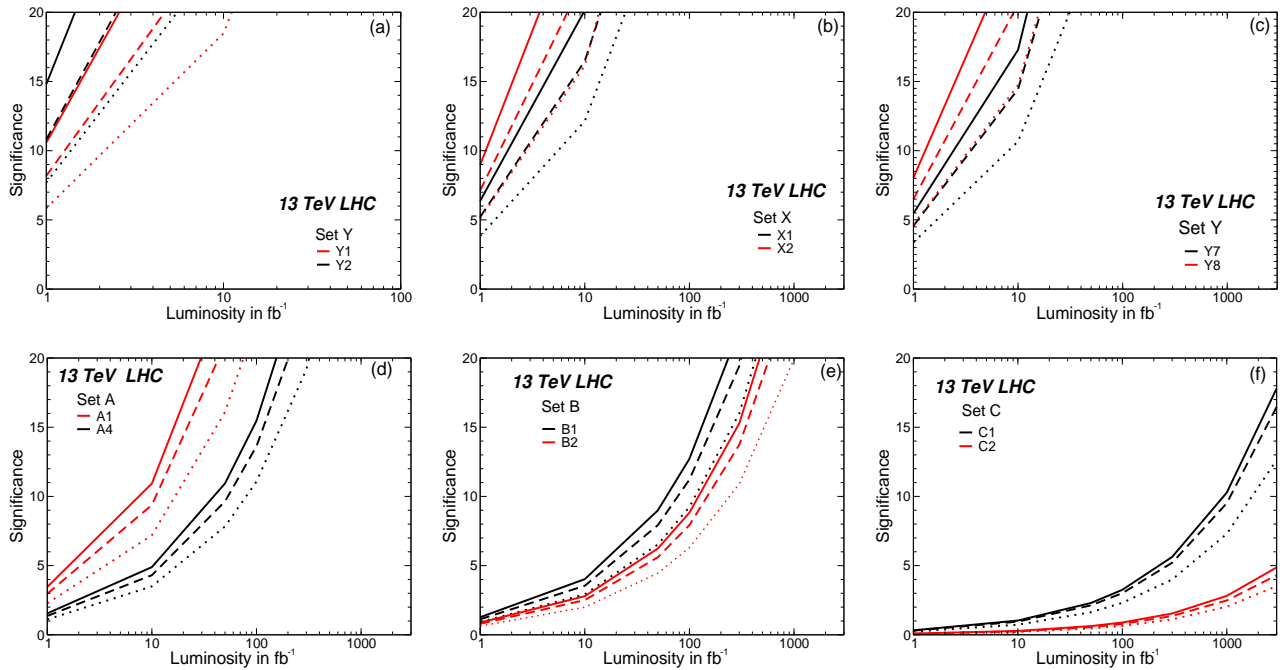

\begin{minipage}[]{0.33\linewidth}
%\subcaption{\label{a}Set Y}
\includegraphics[width=5.3cm,height=4.2cm]{a.eps}
\end{minipage}
\begin{minipage}[]{0.33\linewidth}
%\subcaption{\label{b} Set X}
\includegraphics[width=5.3cm,height=4.2cm]{b.eps}
\end{minipage}
\begin{minipage}[]{0.33\linewidth}
%\subcaption{\label{c} Set Y}
\includegraphics[width=5.3cm,height=4.2cm]{c.eps}
\end{minipage}
\vskip 10pt
\begin{minipage}[]{0.33\linewidth}
%\subcaption{\label{d} Set A}
\includegraphics[width=5.3cm,height=4.2cm]{d.eps}
\end{minipage}
\begin{minipage}[]{0.33\linewidth}
%\subcaption{\label{e} Set B}
\includegraphics[width=5.3cm,height=4.2cm]{e.eps}
\end{minipage}
\begin{minipage}[]{0.33\linewidth}
%\subcaption{\label{f} Set C}
\includegraphics[width=5.3cm,height=4.2cm]{f.eps}
\end{minipage}
\caption{\label{fig:4} {\it The discovery significance ($\Zdis$) as a function of the integrated 
luminosity (${\cal L}_{\rm int}$). Solid, dashed and dotted lines represent 
0\%, 25\% and 50\% uncertainty in the background events respectively. 
Events are selected by {\bf S2}.
The best fit values are $|A_1|=3.8$, $|A_5|= 2.3$, represented by Set X1 in figure (b).}}
\end{figure}

As the analysis of the preceding section shows, it is indeed
possible to exclude much of the parameter space favored by the
resolution of the $B$-anomalies, and even contemplate discovery.
Rather than restrict ourselves to simplistic signal-to-noise
estimations, we consider, instead, a slightly more sophisticated
statistical test. Towards this end, let us define the null
hypothesis as the set of events being composed entirely of the
background (irreducible or instrumental). This is to be tested
against the alternative hypothesis, which includes both background
as well as the sought after signal. To summarize the outcome of such
a search, one quantifies the level of agreement of the observed data
with a given hypothesis by computing the $p$-value. This $p$-value
can be converted into an equivalent significance, $Z_{\rm dis}$, for
a Gaussian distributed variable. The exact formulation is
summarized in the Appendix, with the 5$\sigma$ discovery
significance ($\Zdis=5$) and 95$\%$ CL exclusion limit
($\Zexc=1.645$) being given by Eqs.~(\ref{eq:zdis}) and 
(\ref{eq:zexc}) respectively.

%==================================================
\begin{figure}
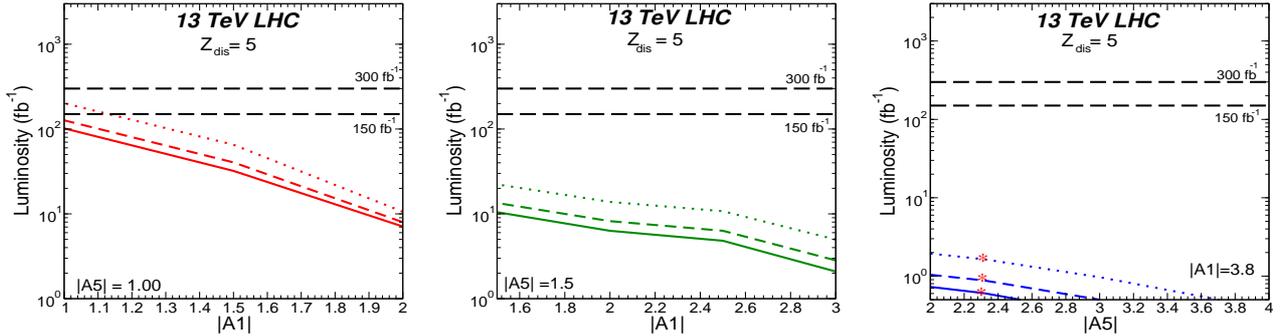

\begin{minipage}[]{0.33\linewidth}
\includegraphics[width=5.3cm,height=4.4cm]{lumi_dis_x.eps}
\end{minipage}
\begin{minipage}[]{0.33\linewidth}
\includegraphics[width=5.3cm,height=4.4cm]{lumi_dis_m.eps}
\end{minipage}
\begin{minipage}[]{0.33\linewidth}
\includegraphics[width=5.3cm,height=4.4cm]{lumi_dis_y.eps}
\end{minipage}
\caption{\label{fig:5} {\it The 5$\sigma$ discovery reach at 13 TeV LHC in 
($\mu^\pm \tau^\mp$) + $b$-jet channel as a function of ${\cal L}_{\rm int}$ and 
$A_1$, $A_5$ (in TeV$^{-2}$). Solid, dashed and dotted lines represent 0\%, 
25\% and 50\% uncertainty in the background events respectively. Events are selected by 
{\bf S2}. The best fit values are $|A_1|=3.8$, $|A_5|=2.3$, represented by red stars.}}
\end{figure}
%==================================================================
We can now consider the discovery and
exclusion prospects of this particular channel under study, 
and compute the required integrated luminosity
(${\cal L}_{\rm int}$) as a function of model parameters $A_{1,5}$.
A straightforward computation,
using the signal and background cross sections calculated in the
last section, leads to very optimistic results though. However, 
the background is not known with a very good precision. To account 
for this, we include a variance $\Delta_b$ in the background 
while calculating the discovery significance and exclusion from 
Refs.~\cite{Cowan:SLAC2012,Cowan:2010js,COUSINS2008480} (see Refs.\cite
{Agarwalla:2018xpc,Kumar:2015tna} for more detail). 
With the current LHC data and the data that
LHC will take in the future, the systematic experimental
uncertainties in the estimation of the SM backgrounds are expected
to be reduced significantly and the detector response is also
expected to be better in future. For example, by comparing
Ref.~\cite{Aaboud:2016tnv} and Ref.~\cite{Aaboud:2017phn}, one can see
find how the systematic uncertainty has reduced so far at LHC. So,
additionally, we also assume in Eqs.~(\ref{eq:zdis}) and
(\ref{eq:zexc}) that a part of the systematic uncertainty falls as
$1/\sqrt{\cal L_{\rm int}}$. 

In Fig.~\ref{fig:4}, we plot the discovery significance ($\Zdis$) for
different values of $A_{1,5}$ (note that the signal cross section and,
hence, the significance is essentially independent of the sign of the
Wilson coefficients) as a function of the integrated luminosity
(${\cal L}_{\rm int}$) for some benchmark points. It is evident from
Fig.~\ref{fig:4}(b) that, for Set X, a small value of ${\cal L}_{\rm 
int}$ is required to achieve 5$\sigma$ discovery significance. This
is because of the large signal cross-section, as can be seen from
Table~\ref{tab:2}. For the best-fit scenario (X1), $Z_{dis}\geq5$ can
be achieved with ${\cal L}_{\rm int}\sim $ 2 fb$^{-1}$, even with a
50\% uncertainty in the background estimation. For a wide range of
values of $A_1$ and $A_5$, $Z_{\rm dis} > 5$ is also achievable with
luminosity achieved so far at the LHC, as can be envisaged from
Fig.~\ref{fig:4}. The significance, as expected, depends on the
signal cross-section, which in turn depends on $A_1$ and $A_5$. Thus,
with lower values of these couplings (such as C3 and C4), one has to
wait for a larger ${\cal L}_{\rm int}$. Overall, we find that a large
region of the model parameter space can be probed with 5$\sigma$ or
higher significance with current LHC data.

In Fig.~\ref{fig:5} we show the variation of the model parameters
$A_1$ and $A_5$ as a function of ${\cal L}_{\rm int}$ that is needed
to get $Z_{\rm dis}=5$. The horizontal black dashed lines represent
the current ${\cal L}_{\rm int}$ at the LHC; 150 fb$^{-1}$ for each of
the two experiments ATLAS and CMS, and 300 fb$^{-1}$ combined. In
Fig.~\ref{fig:5}, we show this variation for Set B (left), Set A
(middle), and Set X (right); the values of $A_1$ and $A_5$ are chosen
in such a way as to satisfy all the low-energy constraints, as
mentioned before. If the values of $A_1$ and $A_5$ lie close to their
best fit values, it is evident from Fig.~\ref{fig:5} (middle and
right) that even a small ${\cal L}_{\rm int}$ is sufficient to either
validate or falsify the model, which should be the case once the
present dataset is fully analyzed. Fig.~\ref{fig:5} (left) shows that
with smaller values of $|A_1|$ and $|A_5|$ (Set B) it is likely to
reach 5$\sigma$ significance with current LHC data, even if the
uncertainty in the background estimation is 50\% or more. For Set C,
much higher luminosity is required for 5$\sigma$ discovery, hence we
refrain from showing the corresponding plots.
%=========================================================
\begin{figure}
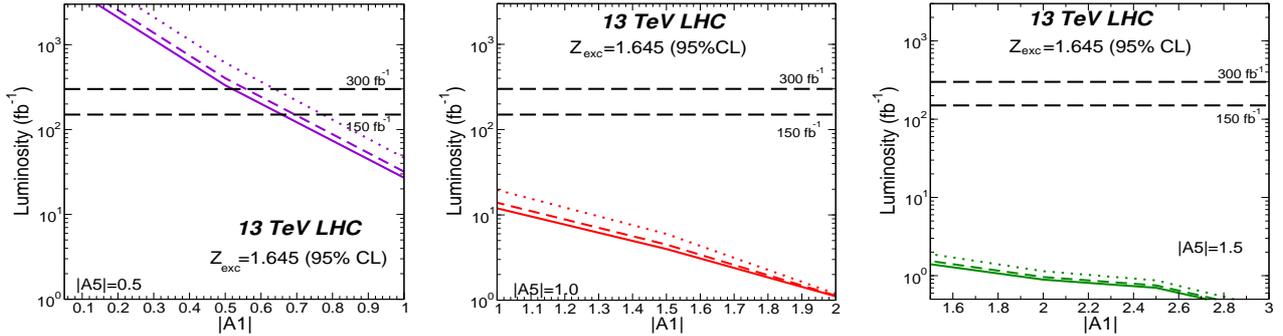

\begin{minipage}[]{0.33\linewidth}
\includegraphics[width=5.3cm,height=4.4cm]{lumi_exc_n.eps}
\end{minipage}
\begin{minipage}[]{0.33\linewidth}
\includegraphics[width=5.3cm,height=4.4cm]{lumi_exc_x.eps}
\end{minipage}
\begin{minipage}[]{0.33\linewidth}
\includegraphics[width=5.3cm,height=4.4cm]{lumi_exc_m.eps}
\end{minipage}
\caption{\label{fig:6} {\it 95$\%$ CL exclusion limits at 13 TeV LHC in the ($\mu^\pm \tau^\mp$) 
+ $b$-jet channel as a function of ${\cal L}_{\rm int}$ and $A_1$, $A_5$ (in TeV$^{-2}$). 
Solid, dashed and dotted lines represent 0\%,25\% and 50\% uncertainty in 
the background events respectively. Events are selected by {\bf S2}. 
}}
\end{figure}
%==============================================

In Fig.~\ref{fig:6}, we show the exclusion limits on $A_1$, keeping
$A_5$ as a parameter, in terms of ${\cal L}_{\rm int}$. The plots,
from left to right, are for Sets C, B, and A respectively. We do not
display the corresponding plots from Sets X and Y as the required
luminosity is quite small, as discussed before. Overall with the
current LHC data 95$\%$ CL exclusion limits can be set in a large
region of the model parameter space including the small values of
$A_1$ and $A_5$.
%======================================================
\section{Outlook}
%======================================================

The not-too-insignificant anomalies in semileptonic $B$ meson decays
point towards some new physics that violate 
lepton flavor universality. One interesting option
is to consider effective dimension-6 operators of the form $a_{ij} \,
(\bar{b}\Gamma_i s) \, (\bar\tau\Gamma_j\tau)$ where $\Gamma_i$ are
operators in the Dirac space. The charged current counterpart of this
operator---arising automatically when $SU(2)_L\otimes U(1)_Y$
symmetry is imposed---may explain the $R(D)$ and $R(D^\ast)$
anomalies, while this operator itself, aided by lepton flavor mixing,
may lead to a possible explanation of the $R_K$ and $R_{K^\ast}$
results. Without the knowledge of the ultimate ultraviolet-complete
theory, the most prudent way to explore the parameter space for new
physics is in terms of the Wilson coefficients. Nominally, these would
go as $\lambda^2/M^2$, where $\lambda$ is some dimensionless coupling,
and $M$ is the mass of the integrated-out mediator, which may be taken
as the scale of new physics.

In this paper, we explored direct signals from such class of operators
at the LHC. While probing the structure of the most general set of
such four-fermi operators would be difficult, the task has been eased
by the analyses of Refs.~\cite{Choudhury:2017qyt,
Choudhury:2017ijp,Bhattacharya:2019eji} which have shown that the
said anomalies can be very satisfactorily resolved in terms of just
two Wilson coefficients, denoted by
$A_{1,5}$. We adopt this simplified structure and also examine how
well these can be explored in terms of the integrated luminosity. As
long as $M$ is much above the scale being probed by the LHC, whether
the new operators are generated through an extra $Z'$, or leptoquarks,
or some other dynamics, is irrelevant.

The signal that we focused upon is an unlike-charged $\mu$-$\tau$
pair associated with a $b$-jet, where the muon comes from the leptonic
decay of one of the daughter $\tau$'s. With suitable cuts, one may
reduce the SM backgrounds for this signal to a very small level, and
thus have a very good detection prospect, even with just the currently
collected data, for values of $A_{1,5}$ preferred by the analyses of
Refs.~\cite{Choudhury:2017qyt,Choudhury:2017ijp,Bhattacharya:2019eji}.
However the values of $A_{1,5}$ are constrained from the
non-observation of Br$(B_s\rightarrow \tau \tau)$, and hence cannot be
chosen arbitrarily. Even with an uncertainty in the estimation of the
background, the situation looks quite optimistic. For example,
notwithstanding the agreement of our background estimation, post {\bf
 S2}, with the experimental results of Ref.\cite{Sirunyan:2018jdk},
let us consider the ramifications of an uncertainty as large as $\sim
50\%$ in the background estimation. Even for the point $(\vert
A_1\vert, \vert A_5\vert) = (1.8,1.5)$, somewhat smaller than the best
fit values of these parameters (and, hence, resulting in a smaller
cross section), can be probed with $5\sigma$ significance at ${\cal
 L}_{\rm int}=20$ fb$^{-1}$. Similarly, with the same ${\cal L}_{\rm
 int}$, and a similar uncertainty in the background, the region of
parameter space defined by ($\vert A_1\vert \geq 1.0$, $\vert A_5\vert
\geq 1.0$) can be excluded at 95\% CL. For even smaller values of
$A_i$, one requires a significantly larger luminosity; for example the
region ($\vert A_1\vert \geq 0.8$, $\vert A_5\vert \geq 0.5$) can be
excluded at 95\% CL with 150 fb$^{-1}$.

The case where both the $\tau$'s decay hadronically is not so clean as
this channel, but will be taken up in a subsequent study. We have not
taken the lepton flavor mixing between $\mu$ and $\tau$ into
account. As has been shown in the literature, the mixing angle is
bound to be small ($\sim 0.02$). While the mixing can directly produce
an unlike-sign $\tau$-$\mu$ pair, the production rate is swamped by
the events where one $\tau$ subsequently decays into a muon. Thus, we
do not envisage that a study of this nature will shed any light on the
mixing angle. This would be better investigated by significantly
improving the measurements of lepton flavor violating decays such as
$B\to K^{(*)} \mu \tau$ or $B_s \to \tau^\pm \mu^\mp$.
%=======================================================
\section{Appendix}
%=======================================================
The significance for discovery in terms of signal events ($s$), background 
events ($b$) and the uncertainty in the background ($\sigmab$) 
is~\cite{Cowan:SLAC2012,Cowan:2010js,COUSINS2008480}, 
%============
\begin{equation}
\Zdis = \left [ 2\left ((s+b) \ln \left [\frac{(s+b)(b+\sigmab^2)}{b^2+(s+b)\sigmab^2}\right ]-\frac {b^2}{\sigmab^2} \ln\left [1+ \frac {\sigmab^2 s}{b(b+\sigmab^2)}\right ]\right)\right]^{1/2} .
\label{eq:zdis}
\end{equation}
%============
If $\Delta_b = 0$,
\begin{equation}
\Zdis=\sqrt{2[(s+b)\ln(1+s/b)-s]}.
\label{eq:zdis1}
\end{equation}
%============
In the above equation, if $b$ is large, then we obtain the well known expression
\begin{equation}
\Zdis = s/\sqrt{b}.
\label{eq:zdis2}
\end{equation}
%============
For discovery reach, $\Zdis \geq 5$ corresponds to 5$\sigma$ discovery 
($p<2.86 \times 10^{-7}$). The exclusion limit at a given confidence 
level (CL) is~\cite{Cowan:SLAC2012,Cowan:2010js,COUSINS2008480}
\begin{equation}
\Zexc=\left [2 \left \{ s-b \ln \left (\frac{b+s+x}{2b} \right ) - \frac{b^2}{\Delta_b^2} \ln \left (\frac{b-s+x}{2b} \right ) \right \} -
(b + s - x) (1 + b/\Delta_b^2) \right ]^{1/2},
\label{eq:zexc}
\end{equation}
%============
where
\begin{equation}
x = \sqrt{(s+b)^2 - 4 s b \Delta_b^2/(b + \Delta_b^2)}.
\end{equation}
%============
In the above equation, if $\Delta_b = 0$,
\begin{equation}
\Zexc = \sqrt{2(s - b \ln(1 + s/b))}.
\end{equation}
%============
For a median expected 95\% CL exclusion
($p = 0.05$), we use $\Zexc \geq 1.645$ for different 
$\Delta_b $. 
%========================================================================================================================================
\section*{Acknowledgments}
D.C. and A.K. are supported by the Science and
Engineering Research Board, India under grants CRG/2018/004889 
and EMR/2016/001306 respectively.
N.K. acknowledges the support from the Indo-French Center for
Promotion of Advanced Research (CEFIPRA Project No. 5404-2) and also Dr. D. S. 
Kothari Postdoctoral scheme (201819-PH/18-19/0013).
N.K. also thanks the organizers of ``DAE BRNS Symposium-2018" for
providing the opportunity to present and discuss the results of this
paper. We also thank the referee for useful comments and suggestions. 
%========================================================================================================================================
\bibliographystyle{JHEP}
\bibliography{bstautau}
\end{document}